\documentstyle[12pt]{article}

\begin{document}
\author{A. Lakshminarayan and  N.L. Balazs \\ {\sl Department of Physics,
State University of New York at Stony Brook}\\
{\sl Stony Brook, N.Y. 11794}}
\date{}
\title{Relaxation and Localization in Interacting  Quantum Maps.}

\maketitle

\newcommand{\newc}{\newcommand}
\newc{\beq}{\begin{equation}}
\newc{\eeq}{\end{equation}}
\newc{\beqa}{\begin{eqnarray}}
\newc{\eeqa}{\end{eqnarray}}
\newc{\longra}{\longrightarrow}
\newc{\longla}{\longleftarrow}
\newc{\ul}{\underline}
\newc{\rarrow}{\rightarrow}
\newc{\br}{\langle}
\newc{\kt}{\rangle}
\newc{\hs}{\hspace}
\newc{\eps}{\epsilon}
\begin{abstract}

	Quantum relaxation is studied in coupled quantum baker's
maps.  The classical systems are exactly solvable Kolmogorov
systems$^{(1)}$, for which the exponential decay to equilibrium is
known.  They model the fundamental processes of transport in
classically chaotic phase space. The quantum systems, in the absence
of global symmetry, show a marked saturation in the level of
transport, as the suppression of diffusion in the quantum kicked
rotor, and eigenfunction localization in the position basis. In the
presence of a global symmetry we study another model that has
classically an identical decay to equilibrium, but quantally shows
resonant transport, no saturation and large fluctuations around
equilibrium. We generalize the quantization to finite multibaker
maps.  As a byproduct we introduce some simple models of quantal
tunneling between classically chaotic regions of phase space.

\vspace{1cm}

\end{abstract}

\newpage

\section{Introduction}
           \hs{.2in} The approach to thermal equilibrium proceeds via
a mixing process in phase space engendered by strong global chaos on
the energy shell. The presence of such strong chaoticity in low
dimensional deterministic dynamical systems is one of the important
discoveries of the modern theory of dynamical systems and has been
reviewed extensively [ See for example Percival in ref.2]. Many
features of classical chaos have been explored with the help of
classical maps, whether as Poincar\'{e} sections of the actual flow in
phase space or as abstract transformations of some manifold.

	The study of quantum systems with a view towards the
manifestations of chaoticity has also been extensive$^{(3,4)}$.  The
concept of low dimensional maps play a central role in the study of
classical dynamical systems. This idea can also be exploited to study
quantum dynamical systems. Such quantum maps {\em have} been
constructed and provide some of the simplest models to study the
quantum manifestations of classical chaos.  The quantum maps studied
either quantize time periodic systems$^{(5)}$, or abstract maps like the
cat map$^{(6)}$ and baker's map$^{(7)}$. In these cases we interpret a quantum
map as an unitary operator which propagates states during one time
step, has the proper classical limit and preserves classical
symmetries. Recently Bogomolny has constructed semiclassical quantum
maps from classical Poincar\'{e} sections$^{(8)}$.  Quantum maps indeed
provide the simplest models to study the quantal manifestations of
classical chaos.

	In this note we use quantum maps to investigate the relaxation
of a quantum system to equilibrium. One of the deterministic classical
systems (a representative of a large class of maps) that we will be
quantizing is strongly chaotic and is isomorphic to a stochastic
Markov chain. This exactly solvable system consisting of three coupled
bakers' maps was studied by Elskens and Kapral$^{(1)}$ to model chemical
rate laws and to study the microscopic origins of the rate constants.
We can think of it as modelling two regions in phase space, that are
chaotic within each other and that are coupled by a mixing mechanism
that can be controlled by a parameter. The rate of relaxation to
global equilibrium would depend on this parameter. We will call such a
map the three bakers' map.

Similar situations arise in the study of classical transport, when
diffusion is impeded by the presence of partial barriers like
cantori$^{(8)}$. A phase point might be trapped in a given region of
phase space for a long time and then ejected to some other region to
again spend a long time there. The presence of this trapping could
significantly affect quantum transport. Such effects have been studied
and reviewed by Bohigas et. al$^{(10)}$ using Random Matrix Theory and the
model of coupled quartic oscillators. Our present interest in these
simple maps is threefold: a) we wish to introduce an interaction
between chaotic maps; b) we want to investigate how far quantum
effects combined with the interaction introduce new time scales, or
modify old ones; c) to create simple maps which admit tunneling
phenomena upon quantization.

The quantization of the bakers' map$^{(7)}$ allows us
to quantize the three  bakers' map in a similar fashion. As a byproduct
of this quantization we get some simple models of tunneling
in quantum systems with chaotic classical limits. We display the
eigenangle splitting in one such  model.

	In Sec.2 we introduce and discuss the classical models
to  be quantized. In Sec.3 we quantize these maps, as well as
quantize two other finite multibaker maps$^{(20)}$. In sec.4 we
present the numerical results and in sec.5 we discuss these results.
We end with a summary in sec.6.

\section{ The Classical Maps}

\subsection{ The Maps}

     \hs{.2in} Since the first map we are quantizing is essentially
that introduced by Elskens and Kapral, we refer to their work$^{(1)}$ for
a very exhaustive and illuminating discussion of these maps.  Here we
provide only the essential details necessary to understand this note.
All the maps below are area preserving and piecewise linear. The
second map is a variant of that studied in ref.1, and we introduce it
due to reasons that will become apparent when we discuss the quantum
maps in Sec.3.

The phase space is the square $[0,1) \times [0,1)$, and is composed of
two regions, or ``cells'', $ A \, [0,1/2) \times [0,1)$ and $B \,
[1/2,1) \times [0,1)$, within which we will  place bakers, so that
there is perfect mixing within regions $A $ and $B$, but there is no exchange
of phase space densities between  them. Then we will introduce a third baker
in the region or ``cell'' $C \, [(1-\alpha)/2, (1+\alpha)/2) \times [0,1)$.
Here $0 \leq \alpha \leq 1 $ is the horizontal width of the overlapping
cell $C$ which provides the mixing amongst the cells $A$ and $B$.
$\alpha$ is the control parameter or the strength  of the ``interaction'',
$\alpha/2$ is the overlap of $C$ in each of the disjoint cells $A$ and
$B$. See fig. 1(c).

The first two maps we will consider below have {\em identical} classical
relaxation behaviours, {\em but their quantal versions are vastly
different} (we will introduce further models in sec.3 that will share
some of these relaxation laws). The first map, $F_{\alpha}^{1}$ is
essentially identical to the construction in ref.1, except for the
details of scaling and choice of labels $p$ and $q$. $F_{\alpha}^{1}$
is written as a composition of three baker's maps, $M_{A}$, $M_{B}$ and
$M_{C}$.  $M_{A}$ acts on region $A$ and it leaves the rest of phase
space unaffected.  If $(q,p)
\in A$, and $M_{A}(q,p)=(q^{\prime}, p^{\prime})$, then

\beq  (q^{\prime}, p^{\prime}) =
       \left\{ \begin{array}{cc}
        (2\,q,\,\, p/2),  & 0 \,\leq\, q  < 1/4\\
         (2\,q -1/2,\, (p+1)/2),  & 1/4 \, \leq q < \, 1.
             \end{array}
            \right.
\eeq

If $(q,p) \in B$ and $M_{B}(q,p)=(q^{\prime}, p^{\prime})$, then;

\beq  (q^{\prime}, p^{\prime})= \left\{ \begin{array}{cc}
         (2\,q -1/2,\,\, p/2),    & 1/2 \, \leq q \,<\, 3/4 \\
          (2 \, q -1,\,\, (p+1)/2),  & 3/4 \, \leq q \, < \,1.
            \end{array}
            \right.
\eeq

If $(q,p) \in C $ and $M_{C}(q,p)=(q^{\prime}, p^{\prime})$, then;

\beq  (q^{\prime}, p^{\prime})= \left\{ \begin{array}{cc}
         (2\,q -(1-\alpha)/2,\,\, p/2),    & (1-\alpha)/2 \, \leq q \,<\, 1/2
\\
          (2 \, q -(1-\alpha)/2,\,\, (p+1)/2),  & 1/2 \,
                     \leq q \, < \,(1+\alpha)/2.
            \end{array}
            \right.
\eeq

\beq
	F_{\alpha}^{1}\,=\, M_{C}\,M_{B}\,M_{A}.
\eeq
We also define  just the mixing within the cells $A$ and $B$,
without the interaction (two rectangular bakers side by side in cells
$A$ and $B$) as $M_{AB}^{1}$,
\beq
	M_{AB}^{1}\,=\, M_{A}\,M_{B}.
\eeq
(Here, and in the rest of the paper superscripts on $M$ and $F$ are
indices and not powers).
Thus the cells $A$ and $B$ have disjoint bakers and the baker in $C$
effects a communication between them. When $\alpha =0$, there is no
interaction. Elskens and Kapral$^{(1)}$ showed that the transformation
$F_{\alpha}^{1}$ had strongly ergodic properties. For rational values
of $\alpha$, they constructed finite Markov partitions$^{(11)}$. They
showed that when $\alpha =1/2$, the transformation $F_{\alpha}^{1}$ is
isomorphic to a Bernoulli shift. The motion is ergodic with respect to
the usual Lebesgue measure, the area.  This implies that the Lyapunov
exponent, the measure of the rate of exponential separation of nearby
points is $\lambda \, = (1\,+\,\alpha) Log\, 2$.  All sufficiently
smooth phase space densities evolve into, the uniform distribution, an
unique equilibrium distribution.

The second map we will study, $F_{\alpha}^{2}$, has identical dynamics in cells
$A$ and $C$; we have bakers defined in them via $M_{A}$ and $M_{C}$
(eqns.(1,3)), but in cell $B$ we reflect the baker of cell  $A$, about
the line $q=1/2$, instead of  translating it.  The
entire dynamics in cells $A$ and $B$, $M_{AB}^{2}$ without the interaction $C$,
is as follows. If  $M_{AB}^{2}(q,p)\,=\,(q^{\prime},p^{\prime})$,
\beq  (q^{\prime}, p^{\prime})= \left\{ \begin{array}{cc}
         (2\,q -1/2,\,\, p/2),    & 0 \, \leq q \,<\, 1/4 \\
          (2 \, q -1/2,\,\, (p+1)/2),  & 1/4 \, \leq q \, < \,3/4\\
	 (2 \, q -1,\,\, p/2), & 3/4\,\leq q \, < 1.
            \end{array}
            \right.
\eeq
The complete transformation, $F_{\alpha}^{2}$, during one time step
is
\beq
	F_{\alpha}^{2}\,=\, M_{C}\,M_{AB}^{2}.
\eeq

\subsection{ The Relaxation Laws}

	\hs{.2in} The classical steady state is described by the
microcanonical ensemble.  The approach to this equilibrium is
evidently governed by the parameter $\alpha$; the larger the value of
$\alpha$ the faster will be the approach to the steady state. The
generally expected exponential relaxation to equilibrium is realized
in the maps $F_{\alpha}^{1,2} \equiv F_{\alpha}$ (We do not write the
superscripts when we do not want to distinguish between the two maps).
We can find the relaxation laws by calculating suitable correlation
functions.  We do not expect there to be qualitative differences for
slightly differing values of the parameter $\alpha$, and we
concentrate on $\alpha\, = \, 1/2, \, 1/4, \, 1/8$.  For these cases
we can find the explicit relaxation behaviours via the Markov matrices
that describe the Markov processes isomorphic to these maps.
Irrational $\alpha$ values do not permit a finite partition, but as
noted in ref.1, we expect the relaxation to be close to that of the
rational approximations of $\alpha$. We do not unduly bother about
this because the very act of quantization imposes the requirement that
$\alpha$ be rational.  We relegate details of the Markov matrices to
Appendix A.

	We consider the correlation between  two densities given
by the characteristic  functions $\chi_{A}$ and $\chi_{B}$, defined as follows,

\beq    \chi_{A,B}(q,p)= \left\{ \begin{array}{cc}
         1\,,    & (q,p) \in \,A,B \\
          0\, ,  &  \mbox{otherwise}.
            \end{array}
            \right.
\eeq
	We then propagate the density $\chi_{A}$ in time and find its overlap
with $\chi_{B}$. This is the transition rate from region $A$ to $B$, for
a density that is initially uniformly distributed in $A$.

\beq
	F_{\alpha}^{t} \, \chi_{A}(q,p) \, = \left\{ \begin{array}{cc}
         1 \, ,    &  F_{\alpha}^{-t}(q,p) \in \, A \\
          0 \, ,  &  \mbox{otherwise}.
            \end{array}
            \right.
\eeq
Here $t$ is the time and takes values over the set of integers. The relaxation
law is described by the correlation function,
\beq
	C_{\alpha}(t)\,=\, \int_{0}^{1}\, \int_{0}^{1}\, \chi_{B}(q,p)
	\, F_{\alpha}^{t}\,\chi_{A}(q,p)\,dq\,dp.
\eeq

	The simplicity of the model of three bakers, allows us to
	exactly evaluate these correlation functions. For $\alpha\,
	=1/2\,,1/4\,,1/8, $ they are,

\beq
C_{1/2}(t)\,=\, (1\,-\,2^{-t})/4,
\eeq
\beq
C_{1/4}(t)\,=\, (2 \,-\, \beta_{1}\,\lambda_{1}^{t}\,-\,
 \beta_{2}\,\lambda_{2}^{t})/8,
\eeq
\beq
C_{1/8}(t)\,=\, (4 \,-\, \beta_{3}\,\lambda_{3}^{t}\,-\,
 \beta_{4}\,\lambda_{4}^{t}\,-\,\beta_{4}^{\ast}\,\lambda_{4}^{\ast \, t}
)/16.
\eeq
The first of these eqns. is given in ref.1, and the others can be
derived from the respective Markov partitions. (We indicate how to do
this in Appendix A.)  The approximate values of the constants are,
$\beta_{1}=1.8944,\,\,
\beta_{2}=2-\beta_{1},\,\,\lambda_{1}= .8090,\,\, \lambda_{2}=
1/2-\lambda_{1},\,\, \beta_{3}=3.8479,\,\,\beta_{4}=.07601\,+i\,
.0113,\,\, \lambda_{3}=.9196,\,\, \lambda_{4}=-.2098\,+i\,.3031.$

We observe that there is an exponential decay, or relaxation to
the uniform distribution. The relaxation is in general a sum of
exponentials with small oscillatory contributions; thus there are
multiple relaxation times. The $\lambda_{i},\,\,i=1,2,3,4$, above are
not the Lyapunov exponents, and are independent characteristics of the
mixing system. During the first time step the slope of the relaxation
curves are approximately $\alpha$, while for a further short linear
regime the slope is approximately $\alpha/2$, thereafter the
relaxation is exponential.

The Markov partitions of the map $F_{\alpha}^{2}$ may be taken to be
the same as that of $F_{\alpha}^{1}$; hence the classical relaxation
laws, eqns.(11-13) are {\em identical}. This of course does not mean
that that all the details of the classical mechanics are identical,
(for example the exact locations of the periodic orbits are
different), but the ``macroscopic'' quantities of the systems, like
entropy, Lyapunov exponents, and correlation functions are identical.
This pair of classical models are interesting, {\em because different
microscopic dynamics generate not only the same equilibrium
quantities, but also the same irreversible macroscopic behaviour}.
Two further models we will be studying are described in sec.3.3.  In
Sec.4 we will be comparing $C_{\alpha}(t)$ with  their quantal
equivalents.

\subsection{ Classical Symmetries }

The map $F_{\alpha}^{1}$ has several symmetries that were not
discussed in ref.1, presumably because they have no direct relevance to
the classical relaxation mechanisms  described above.  However we can
anticipate that these not only will significantly modify quantal
transport, but must also be preserved for a correct quantization.
 Consider first the map $F_{\alpha}^{1}$.  $M_{A}$ and $M_{B}$, or
together $M_{AB}^{1}$ inherit the symmetries of the usual bakers' maps,
defined by them. It is invariant under the following symmetry
operations.

\begin{enumerate}
\item $R_{A}$:\, $p \rarrow 1-p,\,\, q \rarrow 1/2 \, -q,$ for cell $A$.
This is the
reflection symmetry about the center of the rectangular cell $A$.
\item $R_{B}$:\, $p \rarrow 1-p,\,\, q \rarrow 3/2 \, -q,$ for cell $B$.
This is the
reflection symmetry about the center of the rectangular cell $B$.
\item $R$ :\, $p \rarrow 1-p,\, \, q \rarrow 1-q$, for the entire phase space
square. This is the global reflection symmetry about the center of the square.
\item $T$ : \,$ q \rarrow (1/2\,+q)(mod\,1)$, this is the global symmetry of
translation in position by 1/2.
\end{enumerate}

These are all canonical phase space symmetries that must be preserved
by the quantization. They are not independent of each other, as a
translation by 1/2, followed by a reflection about the center of the
square is equivalent to reflection about the individual cells, but it
is convenient to define them as here. The quantization proceeds
piecewise, we will first quantize the ``free'' maps $M_{AB}^{1,2}$,and
then quantize the interaction. Thus we will require the quantal
version of $M_{AB}^{1}$ to preserve all the canonical symmetries
listed above. When we add the interaction $M_{C}$, the symmetries
$R_{A}, R_{B}$, and $T$ break, but due to the reflection symmetry of
the interaction the global $R$ symmetry is preserved.

In the case of the map $F_{\alpha}^{2}$, there is the reflection
symmetry about the individual cells $A$ and $B$, $R_{A}$ and $R_{B}$,
but there is no global $R$ or $T$ symmetries. Instead there is an
anticanonical symmetry $S$ due to spatial reflection about the line
$\,q\,=1/2$, $ (p\,\rarrow p,\,q \,\rarrow 1\,-q).$ There are many
ways in which we can break $R$ and $T$ symmetries, but, as noted
earlier, the advantage of $F_{\alpha}^{2}$  is that it shares the same
relaxation laws as $F_{\alpha}^{1}$.

In the next section we will write down the quantal propagators
corresponding to the classical maps $F_{\alpha}^{1}$, and
$F_{\alpha}^{2}$.  We will also quantize two finite ``multibaker
maps''$^{(20)}$. We relegate some details of the quantization to the
Appendix B.

\section{The Quantum Maps}

\hs{.2in} In this section we will write down the unitary quantal propagators,
or quantum maps. The method of quantizing these maps which have no
Hamiltonian generating them, is facilitated by the quantization
procedures of Balazs and Voros$^{(7)}$, developed for the usual
baker's map. We note that there is no ``canonical'' method to
construct the propagator, but the quantum baker of ref.7, has proved
to be a natural example, and has been studied from several points of
views$^{(12, 13, 14, 15)} $. The semiclassical trace can be written as a
periodic orbit sum$^{(13, 14)}$, and heavy scarring has been noted in its
eigenfunctions$^{(15)}$. The eigenangle statistics show level repulsion
and fall into the universality class of GOE Random matrices.  There
are no eigenangle degeneracies and all eigenangles are irrational
multiples of $2\,\pi$.  It is thus a prototype for the study of the
quantal manifestations of classical chaos.

\subsection{The Propagators}

	\hs{.2in} The classical phase space used above is the
unit square (using units such that the maximum position and
momentum values are unity). One can make the classical phase space
compact, by identifying the opposite edges, turning it thereby
into a torus, and making the originally finite $p$ and $q$ periodic.

The original phase torus can be divided into $N$ phase cells of size
$h$. Thus, according to the original ideas of Planck the dynamical
system has $N$ states and the state space is an $N$ dimensional vector
space. Since the original momentum and position operators are not
periodic operators we replace them with their unitary extensions,
denoted as $V$ and $U$.  The eigenvectors of $U$ are the position
eigenkets $|q_{n}\kt$, and the eigenvectors of $V$ are the momentum
eigenkets $|p_{m}\kt$, with $m,n\,=\,0,1,2,\ldots,N-1$. The
transformation functions between the eigenstates of position and momentum are
given by
\beq  (G_{N})_{nm} \,\equiv \,
\br q_{n}|p_{m}\kt\,=\, \frac{1}{\sqrt{N}} \, e^{\frac{2\pi i}{N}
 (n+1/2)(m+1/2)}.
\eeq

Following Saraceno in ref.15 we have adopted antiperiodic boundary
conditions for the states $|q_{n}\kt$ and $|p_{m}\kt$.  $V$ being
the unitary translation operator in position, we have,
\beq \br q_{m+1}|\,=\, \br q_{m}|\,V,
\eeq
and
\beq
        \br q_{m+N}| \,=\,-\br q_{m}|\,=\,  \br q_{m}|\,V^{N}.
\eeq

This requirement introduces the 1/2 in the phase of the Fourier
transforms of eqn.(14), and facilitates the preservation of classical
phase space symmetries. This essentially places the position and momentum
eigenvalues at half integer, rather than at integer sites.

	The unitary operator corresponding to the classical map
$M_{AB}^{1}$ (just two noninteracting bakers placed side by side in
cells $A$ and $B$) in the position representation is given by,
\beq
	B_{AB}^{1}\,=\, G_{N}^{-1}\left( \begin{array}{cccc}
	G_{N/2}^{L} & 0 & G_{N/2}^{R} & 0 \\
	0  & G_{N/2}^{L}  & 0  & G_{N/2}^{R}
	\end{array}
	\right).
\eeq
Although this looks like  a rectangular matrix it is, in fact,
a square one. The matrices $G_{N/2}^{L}$ and  $G_{N/2}^{R}$
are {\em rectangular} matrices of dimensions $ N/2 \times N/4$.
They are simply the vertical left and right halves of the
square matrix $G_{N/2}$, i.e.,
\beq
 (G_{N/2})_{mn}\,= \, \left\{ \begin{array} {ll}
 (G_{N/2}^{L})_{mn},  & 0\,\leq\, m \,\leq\, N/2 -1, \,\,
0\,\leq\, n \,\leq\, N/4 -1,\\
(G_{N/2}^{R})_{mn},  & 0\,\leq\, m \,\leq\, N/2 -1,\, \,
N/4\,\leq\, n \,\leq\, N/2 -1,
\end{array}
\right.
\eeq
while 0 is the null $N/2 \times N/4$ matrix. Thus the ``free'' propagator
$B_{AB}^{1}$ is a product of two simple matrices, and remarkably
enough it is unitary.  $G_{N}^{-1}$ is an unitary matrix, and the
unitarity of the second matrix term follows from the unitarity of
$G_{N/2}$, for this implies the following;

\beq
G_{N/2}^{L}G_{N/2}^{L\,\dagger} +G_{N/2}^{R}G_{N/2}^{R\,\dagger}\,=
\, I_{N/2},
\eeq
\beq
	G_{N/2}^{L \, \dagger}G_{N/2}^{L}\,=\,
        G_{N/2}^{R \, \dagger}G_{N/2}^{R}\,=\, I_{N/4},
\eeq
\beq
		G_{N/2}^{L \, \dagger}G_{N/2}^{R}\,=\,
        G_{N/2}^{R \, \dagger}G_{N/2}^{L}\,=\, 0_{N/4}.
\eeq

	Here  $I_{N/4}$ and $I_{N/2}$ are the $N/4 \times N/4$
and $N/2 \times N/2$ identity matrices  and  $0_{N/4}$ is the
$N/4 \times N/4$ null matrix.

	The propagator for the map $M_{AB}^{2}$, which is a baker
in cell $A$ and a reflected baker in cell $B$ is given by

\beq
	B_{AB}^{2}\,=\, G_{N}^{-1}\left( \begin{array}{cccc}
	G_{N/2}^{L} & 0 & 0 & G_{N/2}^{R} \\
	0  & G_{N/2}^{L}  & G_{N/2}^{R}  & 0
	\end{array}
	\right),
\eeq
	which again is an unitary matrix. Thus we have the ``free''
propagators $B_{AB}^{1,2}$. A few observations are in order here.
These propagators are not in a {\em block diagonal} form, and thus
they engender the quantal phenomenon of {\em tunneling}. The
{\em classically isolated bakers in cells $A$ and $B$ are not quantally
isolated}. This was {\em forced} upon us because we chose to quantize the
phase space as a {\em whole}. We could have considered the bakers in cells
$A$ and $B$ to act on disjoint Hilbert spaces of dimensions $N/2$.
This would simply give us the in position representation the
propagator
\beq
B_{sc} = \left( \begin{array}{cc}
	B_{N/2} & 0\\
	0 & B_{N/2}
	\end{array}
	\right).
\eeq
Here $B_{N/2}$ is the baker on $N/2$ states, the baker for $N$ states is given
by the matrix of eqn.(67).

This propagator would neglect quantal tunneling, and we can consider
it as a semiclassical propagator. Indeed while quantizing the
interacting baker in cell $C$ we will adopt such a procedure, as we do
not know how to do it otherwise. This will not alter qualitatively the
relaxation behaviour, as we expect such tunneling to be quite small.
In sec.3.3 we will introduce models that do not require such piecewise
quantization. The remarkable fact is that $B_{sc}$ and $B_{AB}^{1}$
which we have constructed from purely dynamical arguments must be near
in all matrix metrics, and should tend towards each other in the limit
of large $N$.

Quantizing the action of $M_{C}$ requires fixing a  portion
of the state space and ``baking'' the rest. We will require $N\alpha$
to be an integer, and we take as the interaction the
unitary matrix

\beq
	B_{C}\,=\, \left( \begin{array}{ccc}
	I_{(1-\alpha)N/2} & 0 & 0\\
	0 & B_{N\alpha} & 0 \\
	0& 0& I_{(1-\alpha)N/2}
	\end{array}
	\right).
\eeq

	Here $B_{N\alpha}$ is the baker transform on $N\alpha$
states, and it provides the interaction between cells $A$, and
$B$. Thus the full propagators quantizing the classical maps $F_{\alpha}^{1}$
and $F_{\alpha}^{2}$, given by eqns.(4) and (7), are

\beq
	U_{\alpha}^{1}\,=\, B_{C} \, B_{AB}^{1},
\eeq
and
\beq
	U_{\alpha}^{2}\,=\, B_{C} \, B_{AB}^{2}.
\eeq

Here the subscript $\alpha$ explicitly shows the parameter dependence,
or the strength of the interaction, and has nothing to do with the
dimension of the matrices, which is $N$. The superscripts distinguish
between the maps $F_{\alpha}^{1}$ and $F_{\alpha}^{2}$.
We recollect that classical relaxation  laws for the values
of $\alpha =1/2,1/4,1/8$ are given by the eqns.(11-13).

\subsection{Quantum Symmetries}

	\hs{.2in} This section verifies that the ``free'' propagators
introduced in the previous section preserve the classical symmetries
discussed in sec.2. The implications of these symmetries on the
spectrum is discussed. As we shall see this generates significant
differences between the classical and quantal results. The classical
map $M_{AB}^{1}$, as we recall, is simply two bakers side by side in
cells $A$ and $B$; we take as its quantum map the unitary matrix
$B_{AB}^{1}$ of eqn.(17). The classical map had the symmetries
$R_{A}$, $R_{B}$, $R$ and $T$ which are reflection about the center of
the cells $A$ and $B$, reflection about the center of the square, and
translation by 1/2. The quantal operator corresponding to $T$ is
simply the unitary translation operator of $N/2$ sites in position
($T_{N}\, = \, V^{N/2}$). This is diagonal in the momentum representation
and is given by
\beq
	(T_{N})_{m\,m^{\prime}} \,=\, e^{i\,\pi (m+1/2)}
	\delta_{m^{\prime} \,m}.
\eeq
	$T_{N}^{2}\,=\,-1$,  due to the antiperiodic boundary conditions
on the states, and hence the eigenvalues of $T_{N}$ are $\pm \,i$.

The operator $B_{AB}^{1}$, evaluated explicitly in the {\em momentum}
representation, is given by
\[
	\br p_{m}|B_{AB}^{1}|p_{m^{\prime}} \kt \,=\,
  \frac{\sqrt{2}\,i}{2N}\, \frac{e^{-i \pi(2m-m^{\prime}+1/2)/2}-1}
{\sin{\pi(2m -m^{\prime}+1/2)/N}}.
\]
 \beq\,\,\,\,\,\,\,(1\,+\,e^{-i\pi(m\,-\,m^{\prime})}).
	\left\{ \begin{array}{ll}
	1, & 0\,\leq\, m \leq N/2-1\\
  	e^{-i\pi (m^{\prime}-1)/2}, & N/2\,\leq m\,\leq\, N-1.
	\end{array}
	\right.
\eeq
Any matrix $A_{m \,m^{\prime}}$ commutes with $T_{N}$ if and only if its
entries are non zero when either both $m$ and $m^{\prime}$ are even,
or both are odd.  It is easily seen from the factor $(1\,+\, e^{-i \pi
(m\,- m^{\prime})}) $ that this is true for $(B_{AB}^{1})_{m\,
m^{\prime}}$.  Hence we have
\beq
     [ B_{AB}^{1}\, , T_{N}]\,=\, 0.
\eeq
	(Here we are being sloppy in  using same symbols for the
abstract operator, and its representation in a particular basis.
Of course when computing the commutator we would take the same basis
for both  operators.) Thus the quantal model has the translation
by 1/2 symmetry.

For the other  symmetries, it would be easier to write the
operators in the position basis. We introduce two functions;
\beq
\frac{1}{\nu(m,n)}\,=\, \sin (\pi(m-2n-1/2)/N),
\eeq
\beq
\frac{1}{\mu(m,n)}\,=\, \cos (\pi(m-2n-1/2)/N).
\eeq

 Then an explicit evaluation of the matrix elements from eqn.(17)
yields
\[
\br q_{n}|B_{AB}^{1}|q_{n^{\prime}}\kt \,=\,
\frac{i\sqrt{2}}{2N}(1-e^{i\pi (n-1/2)}).
\]
\beq
\left\{
\begin{array}{cc}
\nu(n,n^{\prime}), & 0 \,\leq \,n \leq N/4 -1.\\
e^{i \pi (m+ 1/2)}  \mu(n,n^{\prime}), & N/4 \,\leq \, n \leq N/2 -1.\\
\mu(n,n^{\prime}), & N/2\,\leq\,n \leq 3N/4 -1.\\
e^{i \pi (m+ 1/2)}  \nu(n,n^{\prime}), & 3N/4 \,\leq \,n \leq N -1.
             \end{array}
               \right.
\eeq

	The classical $R$ symmetry ($p \rarrow 1-p, \, q \rarrow 1-q$)
is quantally implemented by the operator whose matrix elements in the
position representation are given by
\beq
\br q_{n}|R_{N}|q_{n^{\prime}}\kt \,=\, \delta(n\,+\, n^{\prime}\,+\,1 \,-N).
\eeq
$R_{N}^{2}\,=1$, and thus the eigenvalues of this parity operator
are $\pm 1$. Its commutation with $B_{AB}^{1}$ requires that
\beq
\br q_{N-n-1}|B_{AB}^{1}|q_{N-n^{\prime}-1}\kt \,=\,
\br q_{n}|B_{AB}^{1}|q_{n^{\prime}}\kt,
\eeq
which is verified by a straightforward computation, using eqns.(32).
Thus the  symmetry of reflection about the center of the square is
 preserved by the quantum map.
\beq
     [ B_{AB}^{1}\, , R_{N}]\,=\, 0.
\eeq

The classical symmetry of reflection about the center of each cell can
also be implemented quantally. As we noted earlier this symmetry can
be thought of as a composition of $T_{N}$ and $R_{N}$, and thus is not
an independent symmetry; rather, it is the only canonical symmetry that is
present in the map $M_{AB}^{2}$ and its quantal equivalent, the
matrix operator $B_{AB}^{2}$. Hence we will briefly discuss the
corresponding quantal symmetry.

The required symmetry operator, in the position representation,
must be given, up to a phase factor, by $\delta(n\,+n^{\prime}+1
-N/2)$ if $n^{\prime}\, < N/2$,  and  by $\delta(n\,+n^{\prime}+1
-3N/2)$ if $n^{\prime}\, \geq N/2$. We fix the phases by requiring that
in the {\em momentum } representation the symmetry operator be
$  \delta(m\,+m^{\prime}+1-N)$, again, up to a possible phase factor.
The operator consistent with the above requirements is given
in the position representation by
\beq
\br q_{n}|R_{N}^{\prime}|q_{n^{\prime}}\kt \,=\,
\left \{ \begin{array}{rc}
	\delta(n\,+n^{\prime}\,+1\,-N/2), & 0\,\leq\,
	n^{\prime}\, \leq\, N/2 -1 \\
          - \delta(n\,+\, n^{\prime}\,+1\, -3N/2),& N/2\,\leq
           \, n^{\prime}\, \leq \, 3N/2 -1. \\
	\end{array}
	\right.
\eeq
The classical composition is given quantum mechanically as
\beq
R_{N}^{\prime} \,=\, T_{N}\,R_{N}.
\eeq
Our choice of the overall sign of $R_{N}^{\prime}$ is
consistent with this ordering of the operators $T_{N}$ and $R_{N}$.
The commutation of $B_{AB}^{1}$ with $R_{N}^{\prime}$ is then
immediate. We also note that $T_{N}$ and $R_{N}$ anticommute,

\beq
 R_{N}\, T_{N}\,=\,-T_{N}\,R_{N}
\eeq

We can write all the symmetry operations of the propagator $B_{AB}^{1}$
as the group $g$, where
\beq
g\,=\, (I_{N},\,-I_{N},\, R_{N},\,-R_{N},\, T_{N},\, -T_{N},\,R_{N}T_{N},
\, T_{N}R_{N}).
\eeq
This group is isomorphic to the point group $C_{4v}$, the symmetry
group of a square. There are 5 classes, and hence 5 irreducible
representations of $g$, 4 one dimensional ones and 1 two dimensional
representation.  The four one dimensional representations are ruled
out because $R$ and $T$ anticommute, and hence cannot share
eigenvectors from the same ray. Therefore, the eigenvectors of
$B_{AB}^{1}$ must be {\em exactly} doubly degenerate. The global
symmetries of reflection about the center of the square and
translation by 1/2 were {\em both} needed to produce this degeneracy.

In the presence of translation symmetry alone, the levels would have
been nearly degenerate, displaying tunneling splitting.  These would
 then be the simplest models exhibiting such splitting in the presence
of classical chaos. For example if we had (2/3,1/3) bakers defined in
cells $A$ and $B$, we would have translation symmetry but no
reflection symmetry. If we denote the vertical left and right halves
of the square matrix $G_{2N/3}$ by $G_{2N/3}^{L}$ and $G_{2N/3}^{R}$,
( with  dimensions $2N/3 \times N/3$) and the  vertical left
and right halves of the square matrix $G_{N/3}$ by $G_{N/3}^{L}$ and
$G_{N/3}^{R}$, ( with dimensions $N/3 \times N/6$), the propagator
is given in the position representation by

\beq
	B_{AB}^{3}\,=\, G_{N}^{-1}\left( \begin{array}{cccc}
	G_{2N/3}^{L} & 0  & G_{2N/3}^{R}&0 \\
	0  & G_{N/3}^{L}  & 0 & G_{N/3}^{R}
	\end{array}
	\right).
\eeq

	The above model displays tunneling splitting. The level
splittings due to symmetrical structures in classical phase space have
been called dynamical tunneling splittings$^{(19)}$. The usual tunneling
splitting is due to a potential barrier, while dynamical tunneling is
due to the structure of classical phase space.  This has been studied
in the Henon-Heiles system$^{(19)}$ and the anharmonic quartic oscillator
$^{(10)}$. The splittings are usually observed as being due to classically
stable and symmetric orbits. The splittings may be irregular if the
stable regions are surrounded by chaotic orbits. In the case of two
non-interacting baker's maps in two cells, the splittings are solely
due to classically chaotic and symmetrical structures. Let
$\theta_{i}$ be the eigenangles of $B_{AB}^{3}$ arranged in increasing
order and in units of $2 \pi$. Let $\Delta_{i}\, =\, N
\,(\theta_{i+1} - \theta_{i}), \,\, i=1,3,5,...N-1.$ $\Delta_{i}$ are
the tunneling splittings scaled by the factor $N$.  Fig.(3) shows
$\Delta_{i}$ for the case $N=150$. The splittings are seen to be
highly irregular and some of them are as high as the order of Planck's
constant, in contrast to tunneling splittings due to potential
barriers that are exponentially small in Planck's constant.

We now turn to the symmetries of the propagator $B_{AB}^{2}$, which describes
a baker in cell $A$, and a reflected baker in cell $B$. The propagator
in the position represention  is given by the matrix in eqn.(22).
The explicit matrix elements can be obtained by a straightforward
evaluation of geometric sums, and is given by
\[
\br q_{n}|B_{AB}^{2}|q_{n^{\prime}}\kt \,=\, \frac{i \sqrt{2}}{2N}
(1-e^{i\pi(m-1/2)}).
\]
\beq \left\{ \begin{array}{ll}
	\mu(n,n^{\prime}), & 0\,\leq\,n^{\prime}\,\leq\, N/4-1\\
e^{i\pi(m+1/2)}\nu(n,n^{\prime}), & N/4\,\leq\,n^{\prime}\,\leq\, 3N/4-1\\
	-\mu(n,n^{\prime}), & 3N/4\,\leq\,n^{\prime}\,\leq\, N-1,
	\end{array}
	\right.
\eeq
where $\mu(n,n^{\prime})$ and $\nu(n,n^{\prime})$ are defined in
eqns.(30,31).

We saw above that the symmetry of  reflection about the center of each
individual cell is represented by the operator $R_{N}^{\prime}$.
A direct computation demonstrates that
\beq
 [B_{AB}^{2}\,,\, R_{N}^{\prime}]\,=\,0.
\eeq
Thus this symmetry is quantally preserved. The spatial reflection
about the line $q\,=\,1/2$ is an anticanonical symmetry, and is incorporated
quantally as an anti-unitary operator. Let $|\psi\kt $ be  a vector and
its reflected partner be $S_{N}\,|\psi \kt$; then in the position
representation
we should have
\beq
\br q_{n}|S_{N}|\psi \kt \,=\, \br q_{n}|R_{N}|\psi \kt ^{\ast}.
\eeq
Complex conjugation can be thought of as ensuring the ``wrong'' sign
of the momentum that makes the symmetry an anticanonical one.  Thus
the spatial reflection symmetry is  implemented by
$R_{N}\,K$, where $K$ is the conjugation operator. Spatial reflection
symmetry would then require the operator to satisfy the
condition
\beq
  R_{N}\,B_{AB}^{2 \ast}\,R_{N}\,=\, B_{AB}^{2}
\eeq
which is once more easily verified using eqn.(41). The propagator
$B_{AB}^{2}$, being unitary, has its eigenvalues located on the unit
circle. This symmetry implies that they fall into two classes each of
which is the complex conjugate of the other.  Thus the reflection of
the baker in cell $B$ has a drastic effect on the degeneracies of the
``free'' propagator spectrum. It not only lifts such a degeneracy, but
also ``spreads'' the eigenangles on the unit circle. This is one of
the reasons we have introduced the second map, which, though
classically of the same type as that discussed by Elskens and
Kapral$^{(1)}$, is nevertheless quantally significantly different.
Further on we show per se how this may affect transport, for we are
not so much interested in symmetries, as in the implications it has on
quantum relaxation to an equilibrium.

The complete quantum maps are given by eqns.(25,26), and they include
the interaction due to the overlapping baker in $C$. Since this map
$B_{C}$ given by the propagator of eqn.(24) preserves $R$ symmetry,
the complete propagator $U_{\alpha}^{1}$ also preserves $R$ symmetry.
The symmetry of translation and reflection about the individual cells
for the ``free'' propagator $B_{AB}^{1}$ are broken.  This lifts the
degeneracy of the free propagator. But the persistence of the global
$R$ symmetry implies the existence of extended eigenstates. {\em If
the interaction is small we can expect that the near degeneracies
introduce low frequency, large amplitude, oscillations in the relaxation
process, and that the extended eigenstates contribute to large
transport}.  The complete propagator $U_{\alpha}^{2}$, has no global
or local symmetries, and has no near degeneracies. This contrasts with
the first case, and its implications for quantal transport are studied
in the next section.

It must be noted that the usual baker's map has a time reversal
symmetry. It means that a time reversed partner exists reflected about
the secondary diagonal of the square. Such a time reversal symmetry (
which is a combination of a phase space operation and reversing time),
leads to the quantum baker's map possessing an anti-unitary symmetry.  We
have not been able to verify the existence of the corresponding
symmetry in the quantal free propagators $B_{AB}^{1}$ and
$B_{AB}^{2}$. It is a weak symmetry, but not a complete one.  This,
however, does not affect spectral features like degeneracies or extent
of eigenstates. The propagator $B_{sc}$ (eqn.(23)) would exactly
preserve time reversal symmetry, but we take the free propagator to be
$B_{AB}^{1}$, because this associates one Hilbert space to the entire
torus.

\subsection{Further Generalizations and Models}

\hs{.2in} The models we have studied above can be generalized in
many ways. The individual baker's maps can be generalized to have cuts
that partition the phase cells into unequal parts. One such free
propagator was written above, the model $B^{3}_{AB}$.  Another
possible and more interesting generalization is to include more number
of cells. The quantization of three baker's maps placed side by side
and not interacting can be quantized by methods similar to that used
for the propagator $B^{1}_{AB}$. The interactions can then be added in
the above manner. Thus we have a vast collection of simple models
with, in principle, known classical relaxation laws and
quantizations. The case when there are an infinite number of cells
placed along the $q$ direction with baker's maps defined on them and
the interaction is provided by {\em shifting} the original array by one half
in the $q$ direction was studied as the ``multibaker map'' in
ref.20, classically. The trunctation with a finite number of cells
(finite multibaker map) with periodic boundary conditions can then
be quantized using the methods presented in this paper.

We will give some details on this last possible generalization, {\em
primarily because the interaction does not require piecewise
quantization}. As we have noted earlier the interaction $B_{C}$ was
associated with three independent Hilbert spaces. We will once more
consider the case when the free propagator is $B^{1}_{AB}$ or
$B^{2}_{AB}$. The corresponding classical situation is one in which
there are two non-interacting bakers in cells $A$ and $B$, and the
case when there is a baker's map in cell $A$ and a reflected baker's
map in cell $B$.  Again we will consider periodic boundary conditions
so that the free map is defined on a torus.

The interaction can be prepared in many ways. Consider the following
case. Once more we define two baker's maps in the two cells $A$ and
$B$.  The classical map is described by $M^{1}_{AB}$ (eqn.(5)) and the
quantized version is $B^{1}_{AB}$ (eqn.(17)). We then {\em shift} the
entire map by $\beta$ with respect to the original cells, to the left,
in the $q$ direction. This we take as the interaction. Note that if we
do not shift at all $(\beta\, = \,0)$, there is no interaction, there
is no mixing amongst the cells $A$ and $B$. If we shift by $1/2$ there
is again no interaction between the cells $A$ and $B$. The
quantization of such an interaction is straightforward and is given by
\beq
	  	V^{N \beta} \, B^{1}_{AB} \, V^{-N \beta}
\eeq
where $V$, defined by  eqn.(15), is the position shift operator.
$V$ is diagonal in the {\em momentum } basis and is given by
\beq
	\br p_{m^{\prime}}|V| p_{m} \kt \, = \, e^{2 \pi i (m+1/2)/N}
	\delta_{m \, m^{\prime}}.
\eeq
The complete models with the free propagators and the interactions
can be written as
\beq
	U^{3}_{\beta} \, =\,  	V^{N \beta} \, B^{1}_{AB} \,
	 V^{-N \beta} \, B^{1}_{AB},
\eeq
and
\beq
U^{4}_{\beta} \, =\,  	V^{N \beta} \, B^{1}_{AB} \,
	 V^{-N \beta} \, B^{2}_{AB}.
\eeq

The classical maps are once more isomorphic to stochastic Markov
chains, and their relaxation laws can be worked out. We simply
state without proof that the relaxation law for the map with
$\beta=1/8$ is $C_{1/2}$, the relaxation law for the map with
$\beta=1/16$ is $C_{1/4}$. $C_{1/2}$ and $C_{1/4}$ are given by the
eqns.(11,12). The model $U^{3}_{\beta}$ has translational symmetry by
$1/2$, while $U^{4}_{\beta}$ does not. The spectrum of $U^{3}_{\beta}$
is exactly doubly degenerate, while the spectrum of $U^{4}_{\beta}$ is
free from degeneracies.

We prove the double degeneracy of the operator $U^{3}_{\beta}$, as it
does not follow from any manifest symmetry group (there is no
reflection symmetry about any point, although the free propagator as well
as the interaction have such symmetries). The important identity we note
is the following:
\beq
	R_{N} \, V \, R_{N} \, =\, V^{-1}.
\eeq
Thus we have
\beq
U^{3}_{\beta}\,=\, V^{N\beta} \, B^{1}_{AB}\, R_{N} \, V^{N\beta} \, R_{N}
\, B^{1}_{AB} \, =\, \left( V^{N \beta} \, R_{N} \, B^{1}_{AB} \right)^{2},
\eeq
due to the commutation of $B^{1}_{AB}$ with $R_{N}$ (eqn.35).
The operator $ U^{3}_{\beta}$ can thus be expressed as the square of  a
simpler operator $ B_{0} \equiv  V^{N \beta} \, R_{N} \, B^{1}_{AB} $.
$B_{0} $ does not commute with translations or reflections, but it
anti-commutes with translations by 1/2. Thus making use of eqn.(38), we
get
\beq
	B_{0} \, T_{N} \, =\, - \, T_{N} \, B_{0}.
\eeq
	From this it follows that if $\lambda$ is an eigenvalue of
$B_{0}$, so is $ - \lambda$. Thus the spectrum of $U^{3}_{\beta} \, \equiv
B_{0}^{2}$ is doubly degenerate.

\section{Quantum Relaxation, Numerical results}

\hs{.2in} The quantal equivalent of the classical correlation $C_{\alpha}(t)$
of eqns.(11-13) is the probability of transition from cell $A$ to
cell $B$ as a function of time. Thus we consider the quantity
\beq
C_{\alpha}^{Q}(t)\,=\, \frac{1}{N}Trace(U^{t} P_{A} U^{\dagger t} P_{B})
\,=\, \frac{1}{N} \sum_{n^{\prime}=N/2}^{N-1}
\,  \sum_{n=0}^{N/2-1}\, |\br q_{n^{\prime}}|U^{t}|q_{n}\kt|^{2}.
\eeq
Here we may  take for the unitary operator $U$, any one of   the
propagators $U_{\alpha}^{1,2}$ or $U^{3,4}_{\beta}$. $P_{A}$ and
$P_{B}$ are projectors
of the cells $A$ and $B$, which in the position representation
would have the form
\beq
P_{A}\,=\, \left( \begin{array}{cc}
	I_{N/2}&0\\
	0 & 0
	\end{array}
	\right), \,\,
P_{B}\,=\, \left( \begin{array}{cc}
	0&0\\
	0 & I_{N/2}
	\end{array}
	\right).
\eeq

For short times we expect $C_{\alpha}^{Q}(t)$ to be close to $C_{\alpha}(t)$,
the classical correlation, since initial quantal phase space densities
( constructed of  some coherent states), propagate as if they were classical
phase space densities evolving according to  Liouville's equation. In this
section we  probe the longtime behaviour using the quantal models of the
three interacting baker's map.

We write the eigenvalue problem  of the unitary operator $U$ as
\beq
	U\,|\eps_{j}\kt\,=\, e^{i\,\eps_{j}}\, |\eps_{j}\kt,
\eeq
where the eigenstates are $|\eps_{j}\kt$ and the eigenangles are
$\eps_{j}, \, j=0,1,2,\ldots N-1$.
 We have then the time independent part of the quantum correlation to be
\beq
 \bar{C}_{\alpha}^{Q}\,=\, \frac{1}{N} \sum_{n^{\prime}=N/2}^{N-1}
\,  \sum_{n=0}^{N/2-1}\,\sum_{j=0}^{N-1} |\br q_{n^{\prime}}|\eps_{j}\kt|^{2}
	\,|\br q_{n}|\eps_{j}\kt|^{2}.
\eeq

$\bar{C}_{\alpha}^{Q}$ is also approximately the time average of
$C_{\alpha}^{Q}(t)$, $\br C_{\alpha}^{Q}\kt $ over times longer than
the inverse of the smallest eigenangle spacing. The time averaged
correlation is thus directly dependent on the distribution of the
eigenstates over the state space. If we assume that they are spread
out equally in either regions $A$ and $B$, we would get
$\bar{C}_{\alpha}^{Q}$ to be approximately equal to 1/4, which is the
classical time average. Deviations from this must then have
essentially a quantal origin.

In figs.(4-7) we show the transition probability, $C_{\alpha}^{Q} (t)$
for various values of $\alpha$, and $N$.  In all the figures the solid
line represents the quantum model with global $R$ symmetry (labelled
Quantum 1), $U_{\alpha}^{1}$, while the dotted line represents
relaxation in the quantum model $U_{\alpha}^{2}$ (labelled Quantum 2)
which has no global symmetries.  fig.(4) shows the case of
$\alpha=1/8$, in the propagators.  The dashed curves, shown in all the
figures, are the classical relaxation curves of eqns.(11-13) for both
models. Short time behaviours of the same models are shown in fig.(7)
for emphasis.  Recollect that $N$ is the inverse of Planck's constant,
so that we have probed models in a broad range from strongly quantal
to largely semiclassical.

Several features are at once apparent. The quantal transport for both
models $U_{\alpha}^{1,2}$ are slightly higher than the classical
transport for short times fig.(7), and this difference decreases with
decreasing Planck's constant.  We attribute this to quantal tunneling
between the cells $A$ and $B$.  Note that if we had used the block
diagonal matrix $B_{sc}$, eqn.(23) as our free propagator for
$U_{\alpha}^{1}$, the classical and quantal transport would match {\em
exactly} for the first time step. The tunneling effects are quite
small, and in part justify our neglecting them during the interaction
time step.

Apart from tunneling effects we see that quantal transport due to the
propagator $U_{\alpha}^{2}$, the model with no global symmetry, follows
the classical relaxation curve very closely much beyond the linear
regimes. The quantal relaxation curve follows the classical one, up to
a time when quantal effects manifest themselves as saturation of
transport, with an average significantly lower than the classical
saturation value of 1/4. There are fluctuations about this average
that become smaller when  Planck's constant is decreased.  As we
noted earlier lower average is implied if the eigenstates are not
spread between the cells equally. {\em Thus we are led to the appearance of
``localized''  eigenstates}.  For increasing $N$, or decreasing
Planck's constant, the saturation occurs at a higher value, tending
towards the classical uniform distribution.

Figs.(5,6) show the classical and quantal relaxation curves for the
parameters $\alpha\,=1/4,1/2$.  On increasing the ``interaction
strength'' ,$\alpha$, the average quantal steady state approaches the
classical one, and is implied by a gradual delocalization of the
eigenstates. Such effects have been observed in the quantized standard
map, and we will discuss this below.  The same figures discussed above
also show the relaxation behaviour of the quantum map
$U_{\alpha}^{1}$, corresponding to the classical map
$F_{\alpha}^{1}$.  As noted earlier the map $F_{\alpha}^{1}$ has the
same relaxation laws as $F_{\alpha}^{2}$, but we observe that the
quantal relaxation behaviour of the corresponding map $U_{\alpha}^{1}$
is vastly different from that of $U_{\alpha}^{2}$; there is an
anomalously large transport and large fluctuations about the {\em
classical } average. This is especially apparent for smaller values of
Planck's constant, that is when the quantum effects are fully
operative.  The resonant transport between cells $A$ and $B$ actually
{\em decreases} with Planck's constant till the average becomes
slightly lower than the classical equilibrium, indicating a weak
localization. This is apparent for example in fig.(4) for $N=64$.

In fig.8 we plot the relaxations in the models $U^{3,4}_{1/8}$ and
$U^{3,4}_{1/16}$ for $N=64$. We note that the model $U^{3}_{\beta}$
(Quantum 3, in figs.), with exactly degenerate spectrum has very large
fluctuations about the equilibrium, while the model $U^{4}_{\beta}$
(Quantum 4, in figs.) has small fluctuations and is similar in
behaviour to the model $U^{2}_{\alpha}$, which for comparison we plot
in the same figures. The larger fluctuations in the case of the
degenerate model $U^{3}_{\beta}$ is due to the lack of phase
cancelletations, as there are only $N/2$ different eigenangles.

Recall that the map $U_{\alpha}^{1}$ has the global symmetry of
reflection about the center of the square $R$. The commutation of
$R_{N}$ with $U_{\alpha}^{1}$, implies that the eigenstates are of
either even or odd parity; there are no other symmetries, the
translation symmetry of the free propagator $B_{AB}^{1}$ being broken by
the interaction. Any eigenstate can be written as
\beq
      \left( \begin{array}{c}
		|\psi\kt \\
		\pm R_{N/2}|\psi \kt
		\end{array}
		\right),
\eeq
	and is hence is delocalized, in the sense that they are
are distributed exactly equally (in the position basis) over states that
span cells $A$ and $B$.  This would in part lead to the higher averages.
The short time behaviour of the model $U_{1/8}^{1}$,  fig.(7),
shows  significantly higher transport rate than the classical.

We show several eigenstates in fig.9. Four eigenstates of
$U_{1/8}^{2}$ for $N=32$ are shown in (a)-(f). The localization of
some eigenfunctions to either cell $A$ $( n \leq N/2-1)$ or cell $B$
$(n \,\geq N/2)$ strongly suggests that the localization is due to the
lack of classical diffusion. The $R$ symmetry of each individual cell
is broken by the interaction, but may be weakly present.  Not {\em
all} the eigenstates are thus localized; a significant number of them
are largely delocalized and an example is shown in fig.9(f). This
indicates that the localization we have observed is not
``perturbative''. Indeed we have coupled at least $1/8 $ of the
position states by the interaction.

In this last figure the scarring from the fixed points at the corners
of the square is evident.  The structure of eigenfunctions within
regions of localization is still under study [5]. The quantum baker's
map eigenfunctions were scarred by many classical periodic orbits
$^{(15)}$.  Periodic orbit scarring has been observed in many systems
(see Heller in ref.2), since it was first observed in the stadium
eigenfunctions$^{(16)}$.  Scarring is also a type of localization, and we
can expect that the models we have studied, which are coupled bakers'
maps will display this phenomenon. In these cases the structure of
eigenfunctions within the localized regions need not be random, but
can be very regular. The set of periodic orbits that scar certain
eigenfunctions will then be dependent on features of classical
transport.  Some eigenstates of $U_{1/8}^{1}$ are shown in fig.9(g,h)
and the $R$ symmetry is apparent.  The plots are the moduli of the
eigenstates in the position basis.

\section{Discussions}

\hs{.2in}  We have used  models of three interacting baker's maps
 to show the influence of localization and delocalization on quantal
relaxation to an equilibrium. The delocalization in the eigenstates of
the map $U_{\alpha}^{1}$ is simply due to the global $R$ symmetry.
The localization of the eigenstates of the map $U_{\alpha}^{2}$ was
gradually removed by increasing the parameter ${\alpha}$, which also
effects a larger classical transport. Thus the same parameters affect
naturally the classical rate laws, and quantal eigenstates.

Previous studies that have observed similar effects include the much
studied quantized standard map variously known as the quantum kicked
rotor, or the quantum Chirikov map$^{(5)}$. The classical model is a
kicked rotor, with the time dependent Hamiltonian
\beq
H\,=\, p_{\theta}^{2}/2 \, +\, k\, \sin \theta \sum_{n=-\infty}^{\infty}
	\, \delta(t\,-\,nT).
\eeq
Above a critical value of $K\,=\,kT$, the classical motion becomes
unbounded, and a diffusive growth in momentum, given by
\beq
	\br (p_{\theta}-\bar{p_{\theta}})^{2}\kt \,=\, Dt
\eeq
is observed.

	Here $D$ is the diffusion constant $= K^{2}/2$.  The quantal
behaviour of this time dependent system was observed to exhibit
classical diffusive growth in momentum for short times, and then a
saturation. Thus quantum mechanics is supposed to ``suppress
classical chaos''. This reflects the localization of the eigenstates
in the unperturbed $(k=0)$ basis. Although this model has been mapped
on to a 1D tight binding model of solid state physics$^{(17)}$, the
dynamical origins of localization is not yet fully understood$^{(10)}$.

The models we have begun the study in this paper {\em differ from the
standard map in the two essential and related details; the motion is
always bounded, and the diffusive growth is replaced by a relaxation
to an unique equilibrium state via mixing}.  The standard map on the
torus has also been quantized$^{(5)}$, and it should be interesting to
compare these two models. The classical dynamics of the three bakers
is exactly solvable, and also the dynamics has a simple geometrical
picture. The standard map on the torus on the other hand has more
parameters thus providing us with a range of models.

 The three baker's map models the motion on a single energy surface of
an time independent Hamiltonian system, in which different chaotic
regions are connected by an interaction that depends on some
parameter. Classical transport in the presence of partial barriers
$^{(9)}$, such as cantori presents such a situation.  There is chaotic
mixing within separate regions of phase space and a slow crossing over
between these regions.  In other words a single phase point spends a
long time in each region before crossing over into the next.

Recently Bohigas et. al$^{(10)}$ have studied such transport with the help
of Random Matrix Theories and illustrated it with the example of an
anisotropic coupled quartic oscillator. They introduced the term
``semiclassical localization'', as opposed to the ``dynamical
localization'' in the standard map. The localization in the quantal
three baker's map is presumably connected to the ``semiclassical
localization'', although there is nothing really semiclassical about
them. The models introduced here may provide an ideal testing ground
to further study this phenomenon, as they have no additional
complications, and are finite dimensional quantum systems requiring no
artificial truncation of a basis.

\subsection{ Break Time and Localization}

\hs{.2in} While saturation of relaxation, or suppression of diffusion may be
implied by localization of some eigenfunctions in a particular basis,
the natural question is why this localization at all?  We dwell
briefly and qualitatively on the notion of the break time$^{(18, 2)}$ as a
mechanism that localizes eigenstates, at least in the kind of
situations we have modelled using the three bakers' maps. Classical
mechanics is characterised by the continuity of phase space and the
consequent possible long time exploration of trajectories on fine
scales; while in the quantum mechanics of bounded systems the energy
spectrum is discrete and the evolution is quasiperiodic at best. The
discreteness of the spectrum is resolved after a finite time, the
break time, which is roughly proportional to the inverse average level
density, in this case, eigenangle density. ( we have defined the
eigenangles as in eqn.(54)).  Eigenfunctions can be found as the time
Fourier transform of propagating an initial state, and the effective
exploration of the state occurs within the break time. Thus the
eigenfunctions will be localized if the exploration (in our case
simply in the configuration space) is limited.  For instance in the
three baker's models states localized well away from the line $q=1/2$
have such a possiblity of becoming localized, leading to a localized
eigenstate.

\section{Summary}

\hs{.2in}  We have discussed several models showing simple relaxation to
an equilibrium state, both classically and quantum mechanically.  In
the absence of global symmetries, the relaxation can be significantly
retarded, and even suppressed by quantum effects. In these cases we
find significant localization of the eigenstates. Even in the presence
of global symmetry and in the deep semiclassical regime we find a
small difference between the quantum steady state and the classical
steady state, implying the existence of weak localization. These
simple models of quantal transport in bounded systems thus display a
rich structure.  We have also introduced some simple models of
tunneling between classically disjoint and chaotic regions of phase
space.

\newpage
\begin{center}

{\bf Appendix A}
\end{center}
\hs{.2in} Elskens and Kapral in ref.1 showed that the three bakers'
transformation is isomorphic to a finite Markov shift for rational
values of $\alpha$.  We include the partitions for completeness, and
to demonstrate that the maps $F_{\alpha}^{1,2}$, have identical
partitions and hence identical relaxation laws given by eqns.(11-13).
The details differ from that of ref.1 in notations and scale, and also
in the interchange of position and momentum coordinates. The latter
change means that our forward iterated partitions are the backward
iterated partitions of ref.1, and vice versa.

If $\alpha\,=\,a/b$ for $a$ and $b$ integers, one  partitions the unit
square of the phase space into $2\,b$ vertical bands consisting of
rectangles $[k/2\,b,\,(k+1)/2\,b) \times [0,1)$;  $k$ is an integer
such that $ 0\,\leq k\,\leq 2\,b-1$. For $b\,=2$, this partition is
illustrated in fig.2(a). The rectangles of the partition $Q_{k}$, (also called
``atoms'') are labelled by integers. The forward iterate of
this partition of the phase space square is shown in fig.2(b) for the
map $F_{1/2}^{1}$ and in fig.2(c) for the map $F_{1/2}^{2}$. These
maps were defined in the text by  eqns.(1-7).

The stable and unstable directions of both  maps are globally
parallel to the $p$ and $q$ axis respectively, just as for the
individual bakers map. A forward iterate of the partition sits within
the original partition in such a way that there is no further
partitioning in the stable direction, but there are additional
partitions in the unstable direction. In other words, the unstable
manifolds that form part of the partition in fig.2(a), is a proper
subset of the unstable manifolds that form part of the partition in
fig.2(b). Similar conditions are satisfied by the part of the
partition boundary that is formed by stable manifolds, under backward
iterations.

The forward partitions are labelled by symbols whose first integer
represent the atom they belong to at ``present'' and the second integer
refers to the atom they just came from.  Thus the phase square gets
partitioned horizontally by forward iterates into finer regions. The
backward iterates, not illustrated here, similarly partition phase
space vertically. This is the requirement for a partition to be a
Markov partition$^{(11)}$.

We see from comparing the figs.2(b) and (c) that the partition works for
both the maps $F_{\alpha}^{1,2}$, in an essentially identical way.
A measure, or probability, equal to their area is assigned to the atoms,
$p_{k}\,= \, \mu(Q_{k})$ and we write also $p_{kl}\,=\,\mu(Q_{kl})$.
The transition probability from atom $Q_{l}$ to atom $Q_{k}$, is
$m_{kl}\,=\,\mu(Q_{kl})/\mu(Q_{l})\,=\,2\,b \mu(Q_{kl})$.

The  matrix of transition probabilities for $\alpha =1/2$, obtained from
either figs.2(b) or
(c), is then
\beq
\left( \begin{array}{cccc}
	1/2&1/4&1/4&0\\
	1/2&1/4&1/4&0\\
	0& 1/4&1/4&1/2\\
	0& 1/4&1/4&1/2
	\end{array}
	\right).
\eeq

The case of $\alpha=1/2$ is  also isomorphic to a  finite Bernoulli
shift, as shown in ref.1.  The Markov matrices corresponding to
$\alpha=1/4$ and $1/8$ may be similarly constructed. These are all doubly
stochastic matrices; because the atoms of the partition had equal
measures, they have an unique equilibrium state corresponding to
the uniform distribution.

The correlation function in the text may now be
evaluated. The densities we have chosen are particularly
simple as they can be constructed out of the atoms of the
Markov partition. Then the evaluation of correlations is
equivalent to the problem of finding the powers of the
Markov matrices. This leads to the eqns.(11-13).

\begin{center}
 {\bf Appendix B}
\end{center}
Here we give some details of the quantization of $M_{AB}^{1}$,
which is simply two bakers sitting side by side. Since, naturally, this
depends strongly on the quantization of a lone baker, We refer to
refs.3,7,15 for elaborations and emphasize here the differences
that arise. The lone baker map {\em on the unit square} is given by
the transform,

\beq  (q^{\prime}, p^{\prime}) =
       \left\{ \begin{array}{cc}
        (2\,q,\, p/2),  & 0 \,\leq\, q  < 1/2\\
         (2\,q -1, (p+1)/2),  & 1/2 \, \leq q < \, 1.
             \end{array}
            \right.
\eeq
 Its quantization proceeds by mimicking quantally the classical partition
into left and right vertical rectangles, fig.1(a)  with a partitioning
of the Hilbert space into two  orthogonal subspaces of dimensions $N/2$,
represented by
$L^{Q}$ and $R^{Q}$. If $|\phi^{L}\kt$ is a vector in $L^{Q}$,
and $|\phi^{R}\kt$ is a vector in $R^{Q}$,
 These spaces are specified by requiring
\beq
 \br q_{n}|\phi^{L}\kt \,=\, 0 \,\mbox{if}\,\, n\,\geq N/2.
\eeq
\beq
\br q_{n}|\phi^{R}\kt \,=\, 0 \,\mbox{if} \,\,n\,\leq N/2-1.
\eeq
Thus we require $N$ to be even. Classically the left partition is
stretched in the $q$ direction and contracted in the $p$ direction, so
that it forms the horizontal bottom half of the phase square.  The
right vertical partition is similarly transformed into the upper
horizontal half of the phase square, fig.1(b). Thus the vector space
is also likewise divided into the orthogonal subspaces spanned by the
vectors $\psi^{B}$ and $\psi^{T} $,specified by the following
conditions:
\beq
\br p_{m}|\psi^{B}\kt \,=\, 0 \,\mbox{if} \,\,m\,\geq N/2.
\eeq
\beq
\br p_{m}|\psi^{T}\kt \,=\, 0 \,\mbox{if}\,\, m\,\leq N/2-1.
\eeq
 This dynamics is quantally translated by requiring that each vector
from the subspace spanned by $|\phi^{L}\kt$ be transformed into a
vector in the subspace spanned by the vectors $|\psi^{B}\kt$.  In the
{\em momentum} representation the transform of a $N/2$ component
vector $\phi^{L}$ from the subspace $L^{Q}$ written in the position
representation is given by the vector
\beq  G_{N/2}(\phi^{L}).
\eeq
See ref.7 for the original quantization.
Here $G_{N/2}$ is the $N/2 \times N/2$ Fourier transform defined for
$N$ sites by eqn.(14).

	 Since any vector can be written as a sum of vectors
from the subspaces $L^{Q}$ and $R^{Q}$, the quantal
propagator in the {\em mixed representation} is thus given by the matrix

\beq
	\left( \begin{array}{cc}
	G_{N/2} & 0\\
	0 & G_{N/2}
	\end{array}
	\right).
\eeq
Transforming to the position basis we get the quantum baker's map

\beq
B_{N}\,=\, G_{N}^{-1}\left( \begin{array}{cc}
	G_{N/2} & 0\\
	0 & G_{N/2}
	\end{array}
	\right).
\eeq
	 In the quantum baker's transformation,  outlined above, the partition
of phase space before and after the transformation fell naturally into
subspaces that had exact quantal projectors associated with them, and
could be easily writen in either the position (before the transformation), or
momentum (after the transformation) basis; figs.1(a,b). A similar partition
before the transform is shown in fig.2(a) for the map $M_{AB}^{1}$. The
partitioning of the square into 4 equal squares after the
transformation cannot be implemented quantally. We can, however,
consider the transformation of the partitions 1 and 3 together into
the bottom half of the square, and similarly partitions 2 and 0
together into the top half of the square. This is the origin of
tunneling when we quantize the map $M_{AB}^{1}$. We will see that
this partitioning is sufficient to describe the quantum map.

We again mimick the partitioning of the classical phase space before the
transformation , fig.2(a), by dividing the Hilbert space into 4 orthogonal
vector spaces of dimensions $N/4$ each. Thus we will require $N$
to be a multiple of 4. Let the four orthogonal spaces have representative
vectors $(\phi^{3}, \phi^{2}, \phi^{1}, \phi^{0})$, and we will require
that
\beq
\br q_{n}|\phi^{3}\kt\,=\, 0 \,\,\mbox{if}\,\, n\,\geq\, N/4
\eeq
with similar conditions on the other vectors.

The transformation  of partitions 3 and 1 to the lower horizontal half of
the square is thus once more implemented quantally as a Fourier
transform on $N/2$ sites. Thus we write the transformed vector in {\em
momentum } representation as,
\beq
	G_{N/2}\left( \begin{array}{c}
			\phi^{3}\\
			\phi^{1}
			\end{array}
			\right)
	= ( G_{N/2}^{L} | G_{N/2}^{R}) \left( \begin{array}{c}
			\phi^{3}\\
			\phi^{1}
			\end{array}
			\right).
\eeq

Here the vector to the right of the $G_{N/2}$ matrix is written in the
{\em position} representation. We have split the matrix $G_{N/2}$ into
a right and a left half, because these are the operators that act on
the individual vectors from the subspaces 3 and 1 respectively. They
are {\em rectangular } matrices af dimensions $N/2 \times N/4$. A
similar argument holds for the transformation from 2 and 0 to the top
horizontal half of the phase square. Thus we get the propagator of
$M_{AB}^{1}$ in the mixed representation

\beq
	\left( \begin{array}{cccc}
	G_{N/2}^{L}&0&G_{N/2}^{R}&0\\
	0 & G_{N/2}^{L} & 0 & G_{N/2}^{R}
	\end{array}
	\right).
\eeq

Upon transforming to the position representation via the inverse Fourier
Transform we get the propagator eqn.(17).
 Quantization of the other maps like $M_{AB}^{2}$, follows from
similar arguments.

\vspace{2cm}
\begin{center}
{\bf Acknowledgements}
\end{center}
Both authors express their thanks to the National Science Foundation
for their partial support. A.L. also thanks the Dept. of Physics,
S.U.N.Y, Stony Brook for partial financial support and Dr. J.J.M.
Verbaarschot for useful discussions.

\newpage

{\bf REFERENCES}

\begin{enumerate}
\item { Y. Elskens and R. Kapral}, {\sl J. Stat. Phys.} {\bf 38}:1027 (1985).

\item { M.J. Giannoni, A. Voros and J. Zinn- Justin}, eds.,
{\em Chaos  and Quantum Physics}, Proceedings of Les Houches Summer School,
 Session LII 1989 (Elsevier, Amsterdam, 1990).

\item { L.E. Reichl}, {\em The Transition to Chaos In Conservative
Classical Systems: Quantum Manifestations} (Springer Verlag, 1992).

\item { M.C. Gutzwiller}, {\em Chaos in Classical and Quantum
Mechanics} (Springer Verlag, New York, 1990).

\item { F.M. Izrailev}, {\sl Phys. Rep.} {\bf 196}:299 (1990).

\item { J.H. Hannay and M.V. Berry}, {\sl Physica D} {\bf 1}:267 (1980).

\item { N.L. Balazs and A. Voros}, {\sl Ann. Phys., (N.Y.)}{\bf 190}:1(1989).

\item { E.B. Bogomolny}, {\sl Nonlinearity} {\bf 5}:805 (1992).

\item { R.S. MacKay, J.D. Meiss and I.C. Percival}, {\sl Physica D} {\bf 13}:55
(1984).

\item { O. Bohigas, S. Tomsovic and D. Ullmo}, {\sl Preprint}, To appear
in Phys. Rep.

\item { I.P. Cornfeld, S.V. Formin and Ya.G. Sinai}, {\em Ergodic Theory}
( Springer Verlag, Berlin, 1982).

\item { P.W.O' Connor, S. Tomsovic}, {\sl Ann. Phys., (N.Y.)} {\bf 207}:218
(1991).

\item { P.W.O' Connor, S. Tomsovic and E.J. Heller}, {\sl Physica D} {\bf 55}
:340 (1992).

\item { A.M.O. De Almeida and M. Saraceno}, {\sl Ann. Phys., (N.Y.)}
{\bf 210}:1 (1991).

\item { M. Saraceno}, {\sl Ann. Phys., (N.Y.)} {\bf 199}:37 (1990).

\item { S.W. Mc Donald  and A.N. Kaufman}, {\sl Phys. Rev. Lett.}
 {\bf 42}:1189 (1979).

\item { Shmuel Fishman, D.R. Grampel and R.E. Prange,} {\sl Phys.
Rev. Lett.} {\bf 49}:509 (1982).

\item { E.J. Heller}, {\sl Phys. Rev. A } {\bf 35}:1360 (1987).

\item{ M.J. Davis and E.J. Heller}, {\sl J. Chem. Phys.} {\bf 75}:246
(1981).

\item{ P. Gaspard}, {\sl J. Stat. Phys.} {\bf 68}:673 (1992).

\end{enumerate}

\newpage

\begin{center}
{\sc \bf Figures}
\end{center}

\begin{description}

\item[Figure 1.] a) The bakers map partition before the transform.
Left and Right partitions are marked. b) The partition after one
time step. L goes into B, and R goes into T. c) The ``cells'',
$A$, $B$ and $C$, shown  schematically for the maps $F_{\alpha}^{1,2}$.
The width of the rectangle $C$ is $ \alpha $.

\item[Figure 2.] a) The partition for the maps $F_{1/2}^{1,2}$.
b) The partition after one time step for the map $F_{1/2}^{1}$.
This is the partition into which the partition of
fig. 2(a) has evolved into. c) The partition after one time step
for the map $F_{1/2}^{2}$.

\item[Figure 3.] The tunneling splitting $\Delta_{i}$, for the
model $B^{3}_{AB}$ with $N=150$.

\item[Figure 4.] The rate of transition from cell $A$ to cell $B$.
The initial distribution is uniform over cell $A$. The solid lines
are for the quantum model $F_{\alpha}^{1}$, indicated in figs. as
Quantum 1. The dotted lines are for the quantum model $F_{\alpha}^{2}$,
indicated in figs. as Quantum 2. All the dashed lines are the classical
correlations given by the eqns.(11-13). Shown here are different values
of inverse Planck's constant, N, for the case $\alpha =1/8$.

\item[Figure 5.] Same as figure 3, except the value of $\alpha$ is 1/4.

\item[Figure 6.] Same as figure 3, except the value of $\alpha $ is 1/2.

\item[Figure 7.] Same as figure 3. Shown are the correlations
over a shorter time scale to emphasize short time features.

\item[Figure 8.] Correlations of the models $U^{3,4}_{\beta}$,
(Quantum 3 and 4) compared with the classical relaxation and the
relaxation of the model $U^{2}_{\alpha}$.

\item[Figure 9.] (a-f) are some eigenfunctions of the quantum map 1,
$U_{1/8}^{1}$ for $N=32$. a-e show eigenfunctions largely localized
to either cells $A$ or $B$, while f shows a largely delocalized
eigenfunctions. g and h are two eigenfunctions for the quantum map 2,
$U_{1/8}^{2}$. The global $R$ symmetry is evident.

\end{description}

 \end{document}